\documentclass[12pt,a4paper]{article}
\usepackage{amsfonts,latexsym}
\usepackage{amsmath,amssymb}
\usepackage{graphicx,color}

\renewcommand{\thefootnote}{\fnsymbol{footnote}}

\begin{document}

\vspace{12mm}

\begin{center}
{{{\Large {\bf Black hole stability in Jordan and Einstein frames}}}}\\[10mm]

{Yun Soo Myung\footnote{e-mail address: ysmyung@inje.ac.kr} and Taeyoon Moon\footnote{e-mail address: tymoon@sogang.ac.kr}}\\[8mm]

{Institute of Basic Sciences and Department  of Computer Simulation, Inje University Gimhae 621-749, Korea\\[0pt]}

\end{center}
\vspace{2mm}

\begin{abstract}
We  investigate  the classical stability of Schwarzschild black hole
in Jordan and Einstein frames which are related by the conformal
transformations. For this purpose, we introduce two models of the
Brans-Dicke theory and Brans-Dicke-Weyl gravity in Jordan frame and
two corresponding models in the Einstein frame. The former model is
suitable for studying the massless spin-2 graviton propagating
around the Schwarzschild black hole, while the latter is designed
for the massive spin-2 graviton propagating around the black hole.
It turns out that the black hole (in)stability is independent of the
frame which shows that the two frames are equivalent to each other.

\end{abstract}
\vspace{5mm}


%{\footnotesize ~~~~Keywords: Classical Theories of Gravity, Spacetime Singularities, Black Holes in String Theory}

\vspace{1.5cm}

\hspace{11.5cm}{Typeset Using \LaTeX}
\newpage
\renewcommand{\thefootnote}{\arabic{footnote}}
\setcounter{footnote}{0}

%%%% Introduction %%%%

\section{Introduction}
The Brans-Dicke theory, one of scalar-tensor theories (STT), has a
non-minimally coupled scalar field $\phi$ to gravity in addition to
metric~\cite{BDT}. The original motivation of Brans and Dicke was
the idea of Mach, in which they put Mach idea in general relativity
to describe a varying gravitational constant. In connection with
Einstein gravity, gravitational constant $G$ is related to an
average value of a scalar field   which is not constant. Since the
Brans-Dicke theory was released, two versions of STT are possible:
one version is on the Jordan and the other is on the Einstein frame
which is related to the former by a conformal transformation and a
redefinition of a scalar field.  One may have a non-minimally
coupled scalar in the Jordan frame, while one may have a minimally
coupled scalar in the Einstein frame. However, the issues of lively
debate which are not yet resolved completely include whether the two
versions of STT are equivalent or not in the classical gravity and
cosmology~\cite{Faraoni:1999hp,Vollick:2003ic,Flanagan:2004bz}.
Whereas many authors support the point of view that two frames are
equivalent, others support the opposite viewpoint.

Here we wish to raise this issue on the stability of black
holes~\cite{Regge:1957td,Zeri,Vish}. It was proposed that the
stability of black holes does not depend on the frame because it is
a classical solution which is considered as a ground
state~\cite{Tamaki:2003ah}. Presumably, the ground state is stable
against small perturbations. Usually, a non-minimally coupled scalar
makes the linearized Einstein equation around the black hole
complicated when one compares  to a minimally coupled scalar in the
Einstein frame~\cite{Kwon:1986dw}. Because of this complication,
some authors have made  conformal transformations to find  the
corresponding theory in the Einstein frame where a minimally coupled
scalar appears.

In this work, we show explicitly that the (in)stability of the
Schwarzschild black hole is independent of  choosing  the frame by
introducing the Brans-Dicke-Weyl (BDW) gravity and its conformal
partner of the Einstein-scalar-Weyl (ESW) gravity. Especially, we
focus on showing the instability of massive spin-2 graviton
propagating around the black hole in the two gravity theories.

\section{Brans-Dicke-Weyl gravity in two frames }

Let us first consider the Brans-Dicke-Weyl (BDW) gravity whose
action is given by
\begin{equation} \label{BDW}
S_{\rm BDW}=\frac{1}{16\pi}\int d^4x \sqrt{-g}\Big[\phi
R-\frac{\omega}{\phi}(\partial
\phi)^2-\frac{1}{2m^2}C^{\mu\nu\rho\sigma}C_{\mu\nu\rho\sigma}\Big],
\end{equation}
where the Weyl-squared term is
\begin{equation}
C^{\mu\nu\rho\sigma}C_{\mu\nu\rho\sigma}=2\Big(R^{\mu\nu}R_{\mu\nu}-\frac{1}{3}R^2\Big)+
(R^{\mu\nu\rho\sigma}R_{\mu\nu\rho\sigma}-4R^{\mu\nu}R_{\mu\nu}+R^2).
\end{equation}
Here the quantities in the second parenthesis present the
Gauss-Bonnet term, which could be neglected because it does not
contribute to equation of motion. Also, we use the Planck units of
$c=\hbar=1$ and $m$ is the mass of massive spin-2 graviton. We note
that the Brans-Dicke action is conformally invariant only for
$\omega=-3/2$ under full conformal transformations
\begin{equation}
\hat{g}_{\mu\nu}=\Omega^2g_{\mu\nu},~~\hat{\phi}=\frac{\phi}{\Omega},
\end{equation}
while the Weyl-squared term of $\sqrt{-g}C^2$ is conformally
invariant under  conformal transformations.  The $\omega=-3/2$ BDW
gravity is related  to the conformal massive
gravity~\cite{Faria:2013hxa}. In this case, the author has found
unstable $s$-mode of massive spin-2 graviton~\cite{Myung:2014aia}.
In fact, Brans-Dicke parameter $\omega=-3/2$ gives a border between
a standard scalar field ($\omega>-3/2$) and a ghost of negative
kinetic energy ($\omega<-3/2$).

 From the action (\ref{BDW}),  the Einstein equation is
derived to be
\begin{equation} \label{equa1}
\Big[\phi
G_{\mu\nu}-\frac{\omega}{\phi}\Big(\partial_\mu\phi\partial_\nu\phi-\frac{1}{2}(\partial\phi)^2g_{\mu\nu}\Big)-\Big(\nabla_\mu\nabla_\nu\phi-g_{\mu\nu}\nabla^2\phi\Big)\Big]-\frac{W_{\mu\nu}}{m^2}=0,
\end{equation}
where the Einstein tensor  is given by \begin{equation}
G_{\mu\nu}=R_{\mu\nu}-\frac{1}{2} Rg_{\mu\nu}
\end{equation}
and the Bach tensor $W_{\mu\nu}$  takes the form
\begin{eqnarray} \label{equa2}
W_{\mu\nu}&=& 2 \Big(R_{\mu\rho\nu\sigma}R^{\rho\sigma}-\frac{1}{4}
R^{\rho\sigma}R_{\rho\sigma}g_{\mu\nu}\Big)-\frac{2}{3}
R\Big(R_{\mu\nu}-\frac{1}{4} Rg_{\mu\nu}\Big) \nonumber \\
&+&
\nabla^2R_{\mu\nu}-\frac{1}{6}\nabla^2Rg_{\mu\nu}-\frac{1}{3}\nabla_\mu\nabla_\nu
R.
\end{eqnarray}
Its trace is zero  ($W^\mu~_\mu=0$). In the limit of $m^2 \to
\infty$, one recovers the Brans-Dicke theory.

 On the other hand, the
scalar equation is given by
\begin{equation} \label{scalar-eq}
\nabla^2\phi-\frac{1}{2\phi}(\partial \phi)^2+\frac{1}{2\omega
}R\phi=0.
\end{equation}
 Taking the
trace of (\ref{equa1}) leads to
\begin{equation} \label{ricci}
R=\frac{\omega}{\phi^2}(\partial \phi)^2+\frac{3}{\phi}\nabla^2\phi.
\end{equation}
Plugging (\ref{ricci}) into (\ref{scalar-eq}), one finds a massless
scalar equation for  $\omega\not=-3/2$ as \begin{equation}
\label{fscalar-eq} \Big(1+\frac{3}{2\omega}\Big) \nabla^2 \phi=0 \to
\nabla^2 \phi=0.
\end{equation}
Finally, we arrive at the trace equation
\begin{equation} \label{fricci}
R=\frac{\omega}{\phi^2}(\partial \phi)^2
\end{equation}
and the Einstein equation
\begin{equation} \label{fequa1}
\Big[\phi
R_{\mu\nu}-\frac{\omega}{\phi}\partial_\mu\phi\partial_\nu\phi-\nabla_\mu\nabla_\nu\phi\Big]-\frac{W_{\mu\nu}}{m^2}=0.
\end{equation}
  Considering the background ansatz
\begin{equation}
\bar{R}_{\mu\nu}=0,~~\bar{R}=0,~~\bar{\phi}= {\rm const},
\end{equation}
Eqs. (\ref{fequa1}) and (\ref{fscalar-eq}) together with
(\ref{fricci})  provide the Schwarzschild black hole solution
\begin{equation} \label{schw} ds^2_{\rm S}=\bar{g}_{\mu\nu}dx^\mu
dx^\nu=-f(r)dt^2+\frac{dr^2}{f(r)}+r^2d\Omega^2_2
\end{equation}
with the metric function \begin{equation} \label{num}
f(r)=1-\frac{r_0}{r}.
\end{equation}
It is easy to show that the Schwarzschild  black hole (\ref{schw})
is also the solution to the Brans-Dicke theory.

Now we transform the BDW action (\ref{BDW}) into the corresponding
action in the Einstein frame by
choosing~\cite{Faraoni:1999hp,Farajollahi:2010ni,Romero:2012hs}
\begin{equation}
\hat{g}_{\mu\nu}=\phi
g_{\mu\nu},~~\hat{C}^{\mu}_{~\nu\rho\sigma}=C^{\mu}_{~\nu\rho\sigma}
\end{equation}
and the scalar field redefinition
\begin{equation}
\phi \to \hat{\phi}=\sqrt{2\omega+3}\ln \phi. \end{equation} Then,
the action of $\omega>-3/2$ BDW gravity in the Jordan frame is
conformally equivalent to the (minimally coupled)
Einstein-scalar-Weyl (ESW) gravity in the Einstein
frame~\cite{Faraoni:1999hp}
\begin{eqnarray}\hat{S}_{\rm BDW}=\frac{1}{16 \pi}\int d^4 x\sqrt{-\hat{g}}
\Big[\hat{R}-\frac{1}{2}
\hat{g}^{\mu\nu}\partial_\mu\hat{\phi}\partial_\nu\hat{\phi}-\frac{1}{2m^2}\hat{C}^{\mu\nu\rho\sigma}\hat{C}_{\mu\nu\rho\sigma}\Big].
\label{Einact}
\end{eqnarray}
Its Einstein equation takes the form
\begin{equation}\label{efein-eq}
\hat{G}_{\mu\nu}-\frac{1}{2}\Big[\partial_\mu\hat{\phi}\partial_\nu\hat{\phi}-\frac{1}{2}(\partial\hat{\phi})^2g_{\mu\nu}\Big]-\frac{1}{m^2}\hat{W}_{\mu\nu}=0
\end{equation}
and the scalar equation is given by
\begin{equation}\label{es}
\nabla^2\hat{\phi}=0. \end{equation} Tracing (\ref{efein-eq}) leads
to
\begin{equation}\label{er}
\hat{R}=\frac{1}{2}(\partial\hat{\phi})^2.
\end{equation}
For $\hat{\phi}=\bar{\hat{\phi}}=$const, one has
$\bar{\hat{R}}_{\mu\nu}=0$ and $\bar{\hat{R}}=0$. This implies that
the Schwarzschild metric (\ref{schw}) is a solution to the Eq.
(\ref{efein-eq}).

\section{Black hole (in)stability in the Einstein frame}

We briefly describe  the stability analysis of the Schwarzschild
black hole found from the ESW gravity in the Einstein frame. For
this purpose, we introduce the perturbations around the black hole
\begin{equation}
\hat{g}_{\mu\nu}=\bar{g}_{\mu\nu}+\hat{h}_{\mu\nu},~~\hat{\phi}=\bar{\hat{\phi}}+\hat{\varphi}.
\end{equation}
Then, Eq. (\ref{er}) yields the non-propagation of the linearized
Ricci scalar as
\begin{equation} \label{eflins-eq}
\delta \hat{R}=0.
\end{equation}
Taking into account  $\delta \hat{R}=0$,  the linearized Einstein
equation (\ref{efein-eq}) is given by
\begin{equation} \label{eflin-eq}
\delta \hat{R} _{\mu\nu}=\frac{1}{m^2}\Big[ \bar{\nabla}^2\delta
\hat{R}_{\mu\nu}+2\bar{R}_{\rho\mu\sigma\nu}\delta
\hat{R}^{\rho\sigma}\Big].
\end{equation}
If one uses the transverse-traceless gauge of $\bar{\nabla}^\mu
\hat{h}_{\mu\nu}=0$ and $\hat{h}=0$ to obtain
 $\delta
\hat{R}_{\mu\nu}=\frac{1}{2}\Delta_{\rm L} \hat{h}_{\mu\nu}$ with
the Lichnerowicz operator $\Delta_{\rm L}
\hat{h}_{\mu\nu}=-\bar{\nabla}^2\hat{h}_{\mu\nu}-2\bar{R}_{\rho\mu\sigma\nu}\hat{h}^{\rho\sigma}$,
Eq.(\ref{eflin-eq}) could be expressed as a fourth-order equation
\begin{equation} \label{fourth-eq}
\Delta_{\rm L}(\Delta_{\rm L}+m^2)\hat{h}_{\mu\nu}=0,
\end{equation}
which may imply two second-order equations
\begin{eqnarray} \label{th-eq1}
&&\Delta_{\rm L} \hat{h}_{\mu\nu}=0,\\
\label{th-eq2}&&(\Delta_{\rm L} +m^2)\hat{h}_{\mu\nu}=0.
\end{eqnarray}
Actually, Eq. (\ref{th-eq2}) corresponds
 to the
massive graviton equation for $\hat{h}_{\mu\nu}$
\begin{equation} \label{hflin-eq}
\hat{h} _{\mu\nu}=\frac{1}{m^2}\Big[ \bar{\nabla}^2
\hat{h}_{\mu\nu}+2\bar{R}_{\rho\mu\sigma\nu}
\hat{h}^{\rho\sigma}\Big].
\end{equation}
We note that although two Eqs. (\ref{eflin-eq}) and (\ref{hflin-eq})
take the same form, but they have different physical natures. The
former equation is a second-order equation for the linearized Ricci
tensor $\delta\hat{R}_{\mu\nu}$, whereas the latter is a suggesting
second-order equation from the fourth-order equation which gives
rise to ghost-like massive graviton for the metric perturbation
$\hat{h}_{\mu\nu}$. Thus, one argues that Eq. (\ref{hflin-eq}) by
itself does not represent a correct linearized equation for studying
the stability of the black hole in the fourth-order gravity.
Importantly, if one uses (\ref{eflin-eq}) instead of
(\ref{hflin-eq}), one might avoid the ghost issue because
(\ref{eflin-eq}) is a genuine second-order equation. This is the
reason why we choose the Ricci tensor perturbation in the study of
the black hole perturbation in the fourth-order gravity. However, we
remark that the Ricci tensor perturbation $\delta\hat{R}_{\mu\nu}$
is not a massive graviton itself but a boosted-up tensor of the
massive graviton \cite{Bergshoeff:2012yz,Bergshoeff:2013vra}. Here
``boosted-up'' means ``boosting up the number of derivatives'',
which indicates just $\delta
\hat{R}_{\mu\nu}(h)=\frac{1}{2}\Delta_{\rm L} \hat{h}_{\mu\nu}$. We
note that hereafter, the massive graviton defined in the Ricci
tensor formalism implies the (boosted-up) massive graviton.

On the other hand, the linearized scalar equation for (\ref{es}) is
\begin{equation}\label{linhsca}
\bar{\nabla}^2 \hat{\varphi}=0, \end{equation} which is surely a
massless scalar equation propagating on the Schwarzschild black
hole. It turned out that the scalar mode does
 not have any unstable modes
 ~\cite{Kwon:1986dw,Myung:2011ih,Moon:2011fw}.
 Explicitly, introducing the tortoise coordinate
$r^*=r+r_0\ln[r/r_0-1]$ and the scalar
 perturbation
 \begin{equation}
\hat\varphi(t,r,\Theta,\Phi)=e^{-ik t}\frac{\hat\psi(r)}{r}
 Y_{lm}(\Theta,\Phi),
\end{equation}
 the linearized
equation (\ref{linhsca}) reduces to the Schr\"odinger-type equation
as
\begin{equation}
\frac{d^2\hat\psi}{dr^{*2}}+(k^2-V_{\hat\psi})\hat\psi=0
\end{equation}
with the potential
\begin{equation}
V_{\hat\psi}=\Big(1-\frac{r_0}{r}\Big)\Big[\frac{l(l+1)}{r^2}+\frac{r_0}{r^3}\Big].
\end{equation}
The potential $V_{\hat\psi}$  is always positive exterior the event
horizon $r=r_0$ for $l\ge 0$, implying that the black hole is stable
against the  scalar perturbation. It is well known  that the
Schwarzschild black hole is stable \cite{Regge:1957td,Zeri,Vish}
against the odd-and even-perturbations with the same potentials in
Einstein gravity because its linearized Einstein equation
(\ref{eflin-eq}) is given by
\begin{equation}\label{cur1} \delta \hat{R}_{\mu\nu}(\hat{h})=0
\end{equation}
in the limit of $m^2 \to \infty$.

We mention that counting  the number of DOF, it might be helpful to
explain intuitively why the Schwarzschild black hole is physically
stable in the Einstein gravity of $m^2 \to
\infty$~\cite{Regge:1957td,Zeri,Vish}, whereas the Schwarzschild
black hole can be unstable in the ESW gravity. We wish to point out
that $h_{\mu\nu}$ is used to describe a massless spin-2 graviton,
while $\delta \hat{R}_{\mu\nu}$ can be taken to describe massive
spin-2 graviton to avoid ghost states. The number of DOF for the
massless spin-2 graviton $h_{\mu\nu}$ is 2 in the Einstein gravity
($m^2\to\infty$ ESW gravity), since one requires $-3$ further for a
residual diffeomorphism after a gauge-fixing. We know that these $2$
DOF correspond to the transverse modes. On the other hand, from Eqs.
(\ref{eflin-eq}) and (\ref{eflins-eq}) together with the linearized
Bianchi identity ($\bar{\nabla}^\mu\delta \hat{R}_{\mu\nu}=0$), the
number of DOF for massive spin-2 graviton $\delta \hat{R}_{\mu\nu}$
in the ESW gravity~\cite{Myung:2013doa,Myung:2013bow} is $10-5=5$,
which includes the longitudinal (would be unstable) modes
\cite{Eardley:1973br,Eardley:1974nw} as well as transverse modes.

The $s$-mode analysis is suitable for investigating the massive
graviton propagation in the ESW gravity, but not for studying the
massless graviton propagation  in the Einstein gravity. In general,
the $s$-mode analysis of the massive graviton with $5$ DOF shows the
Gregory-Laflamme instability~\cite{Gregory:1993vy} which never
appears in the massless spin-2
analysis~\cite{Babichev:2013una,Brito:2013wya}.  The even-parity
metric perturbation is used to define  a $s(l=0)$-mode analysis in
the ESW gravity and whose form is given by $\delta
\hat{R}_{tt},~\delta \hat{R}_{tr},~\delta \hat{R}_{rr}$ and $\delta
\hat{R}_{\Theta\Theta}$ as~\cite{Myung:2013doa}
\begin{eqnarray}
\delta \hat{R}_{\mu\nu}=e^{\Omega t} \left(
\begin{array}{cccc}
\delta \hat{R}_{tt}(r) & \delta \hat{R}_{tr}(r) & 0 & 0 \cr \delta
\hat{R}_{tr}(r) & \delta \hat{R}_{rr}(r) & 0 & 0 \cr 0 & 0 &  \delta
\hat{R}_{\Theta\Theta}(r) & 0 \cr 0 & 0 & 0 & \sin^2\Theta\delta
\hat{R}_{\Theta\Theta}(r)
\end{array}
\right). \label{evenp}
\end{eqnarray}
 Even though one starts with 4 DOF, they are
related to each other when one uses the transverse-traceless
condition of $\bar{\nabla}^\mu \delta \hat{R}_{\mu\nu}=0$ and
$\delta \hat{R}=0$. Hence, we obtain one decoupled equation for
$\delta \hat{R}_{tr}$ from the massive graviton equation. Since
Eq.(\ref{eflin-eq}) is the same linearized equation for
four-dimensional metric perturbation around five-dimensional black
string as~\cite{Babichev:2013una,Brito:2013wya}
\begin{eqnarray} \label{h-eq}
\bar{\nabla}^2h_{\mu\nu}+2\bar{R}_{\rho\mu\sigma\nu}h^{\rho\sigma}=m^2h_{\mu\nu},~~\bar{\nabla}^\mu
h_{\mu\nu}=0,~~h=0,
\end{eqnarray}
we use the GL instability analysis to reveal unstable
modes~\cite{Gregory:1993vy}. Actually,  Eq. (\ref{eflin-eq}) is
considered as a boosted-up version of the massive graviton equation
(\ref{h-eq})~\cite{Bergshoeff:2012yz,Bergshoeff:2013vra}. In
addition, the massive spin-2 polarizations could be  described by
the linearized Ricci tensor
well~\cite{Eardley:1973br,Eardley:1974nw,Newman:1961qr,Moon:2011gg}.
We stress to note that taking the linearized Ricci tensor is the
only prescription to avoid ghosts because the linearized equation
(\ref{eflin-eq}) becomes a fourth-order differential equation when
it is expressed in terms of the metric perturbation $h_{\mu\nu}$.

Eliminating all but $\delta \hat{R}_{tr}$, Eq.(\ref{eflin-eq})
reduces to a second-order radial equation for $\delta \hat{R}_{tr}$
\begin{equation} \label{secondG-eq} A
\delta \hat{R}_{tr}^{''} +B\delta \hat{R}_{tr}^{'}+C\delta
\hat{R}_{tr}=0,
\end{equation}
where $A,B$ and $C$ are given by
\begin{eqnarray}
A~=~-m^2
f-\Omega^2+\frac{f^{'2}}{4}-\frac{ff^{''}}{2}-\frac{ff^{'}}{r},
\end{eqnarray}
\begin{eqnarray}
B~=~-2m^2f^{'}-\frac{3f^{'}f^{''}}{2}-\frac{3\Omega^2f^{'}}{f}+\frac{3f^{'3}}{4f}+\frac{2m^2f}{r}+\frac{2\Omega^2}{r}+\frac{3f^{'2}}{2r}
+\frac{ff^{''}}{r} -\frac{2ff^{'}}{r^2},
\end{eqnarray}
\begin{eqnarray}
C&=&m^4+\frac{\Omega^4}{f^2}+\frac{2m^2\Omega^2}{f}-\frac{5\Omega^2f^{'2}}{4f^2}+\frac{m^2f^{'2}}{4f}+\frac{f^{'4}}{4f^2}-\frac{m^2f^{''}}{2}-\frac{\Omega^2f^{''}}{2f}-\frac{f^{'2}f^{''}}{4f}-\frac{f^{''2}}{2}
\nonumber\\
&&-\frac{2m^2f^{'}}{r}-\frac{\Omega^2f^{'}}{r f}+\frac{f^{'3}}{r
f}-\frac{3f^{'}f^{''}}{r}
+\frac{2\Omega^2}{r^2}+\frac{2m^2f}{r^2}-\frac{5f^{'2}}{2r^2}+\frac{ff^{''}}{r^2}+\frac{2ff^{'}}{r^3}
\end{eqnarray}
with the metric function $f=1-r_0/r$ (\ref{num}). It is worth noting
that the $s$-mode perturbation is described by single DOF but not 5
DOF. The boundary conditions are that $\delta \hat{R}_{tr}$ should
be regular on the future horizon and vanishing at infinity.

Now we are in a position to solve (\ref{secondG-eq}) numerically and
find unstable modes. See Fig. 1 that is  generated from the
numerical analysis. From the observation of Fig. 1 with ${\cal
O}(1)\simeq 0.86$, we find unstable modes~\cite{Babichev:2013una}
for the small Schwarzschild black hole
\begin{equation} \label{unst-con}
0<m<\frac{{\cal O}(1)}{r_0} \end{equation} with  mass $m$.
\begin{figure*}[t!]
   \centering
   \includegraphics{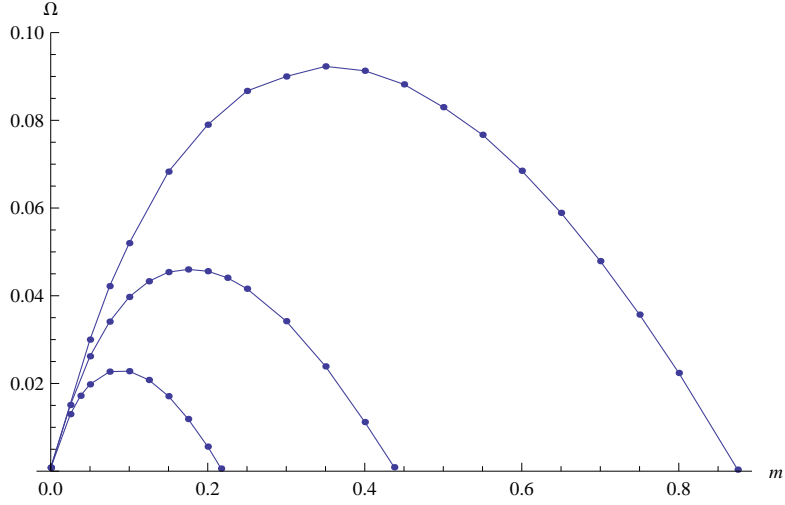}
\caption{Plots of unstable modes on three curves with $r_0=1,2,4$.
The $y(x)$-axis denote $\Omega(m)$. The smallest curve represents
$r_0=4$, the medium denotes $r_0=2$, and the largest one shows
$r_0=1$.}
\end{figure*}
As a consequence, this shows that the region of instability becomes
progressively smaller, as the horizon size $r_0$ increases.

\section{Black hole (in)stability  in the Jordan frame}

We now turn to performing the stability analysis of the
Schwarzschild black hole (\ref{schw}) in the Jordan frame.   To this
end, we introduce the metric and scalar perturbations around the
black hole
\begin{eqnarray} \label{m-p}
g_{\mu\nu}=\bar{g}_{\mu\nu}+h_{\mu\nu},~~\phi=\bar{\phi}(1+\varphi).
\end{eqnarray}
Then, the linearized Einstein equation (\ref{fequa1}) takes the form
\begin{eqnarray} \label{lin-eq}
&&m^2\bar{\phi}\Big[\delta R
_{\mu\nu}-\bar{\nabla}_\mu\bar{\nabla}_\nu\varphi\Big]
\\ \nonumber
&& =\Big[\bar{\nabla}^2\delta
G_{\mu\nu}+2\bar{R}_{\rho\mu\sigma\nu}\delta G^{\rho\sigma}\Big]
+\frac{1}{3}\Big[\bar{g}_{\mu\nu}\bar{\nabla}^2-\bar{\nabla}_\mu\bar{\nabla}_\nu
\Big] \delta R,
\end{eqnarray}
where the linearized Einstein tensor, Ricci tensor, and Ricci scalar
are given by
\begin{eqnarray}
\delta G_{\mu\nu}&=&\delta R_{\mu\nu}-\frac{1}{2} \delta
R\bar{g}_{\mu\nu},
\label{ein-t} \\
\delta
R_{\mu\nu}&=&\frac{1}{2}\Big(\bar{\nabla}^{\rho}\bar{\nabla}_{\mu}h_{\nu\rho}+
\bar{\nabla}^{\rho}\bar{\nabla}_{\nu}h_{\mu\rho}-\bar{\nabla}^2h_{\mu\nu}-\bar{\nabla}_{\mu}
\bar{\nabla}_{\nu}h\Big), \label{ricc-t} \\
\delta R&=& \bar{g}^{\mu\nu}\delta R_{\mu\nu}= \bar{\nabla}^\mu
\bar{\nabla}^\nu h_{\mu\nu}-\bar{\nabla}^2 h \label{Ricc-s}
\end{eqnarray}
with $h=h^\rho~_\rho$. From (\ref{fricci}), we obtain the
non-propagation of the linearized Ricci scalar \begin{equation}
\label{lriccicon} \delta R=0.
\end{equation}
Substituting (\ref{lriccicon}) into (\ref{lin-eq}), one finds the
linearized Ricci tensor  equation
\begin{eqnarray} \label{silin-eq}
\bar{\phi}\Big[\delta R
_{\mu\nu}-\bar{\nabla}_\mu\bar{\nabla}_\nu\varphi\Big]
 =\frac{1}{m^2}\Big[\bar{\nabla}^2\delta
R_{\mu\nu}+2\bar{R}_{\rho\mu\sigma\nu}\delta
R^{\rho\sigma}\Big]=-\frac{1}{m^2}\Delta_L\delta R_{\mu\nu},
\end{eqnarray}
where the Lichnerowicz operator $\Delta_L$ acting on the
transverse-traceless tensor $\delta R_{\mu\nu}$ is introduced to
have the simplicity.  From (\ref{fscalar-eq}), we derive the
linearized scalar equation
\begin{equation} \label{linsca}
\bar{\nabla}^2\varphi=0,
\end{equation}
which implies, as shown in the previous section, that the black hole
is stable against the scalar perturbation.  Here we  note  that
$\delta R_{\mu\nu}$ is taken to describe massive spin-2 graviton to
avoid ghost states, while $h_{\mu\nu}$ is used to describe a
massless spin-2 graviton.

Now, we mention briefly the stability of the black hole in the limit
of $m^2 \to \infty$, reducing to the stability of the black hole in
the Brans-Dicke theory. In addition to the massless scalar equation
(\ref{linsca}), the linearized equation (\ref{silin-eq}) is taken to
be~\cite{Kwon:1986dw,Myung:2011ih,Moon:2011fw}
\begin{equation} \label{jordan-eq}
\delta R_{\mu\nu}(h)-\bar{\nabla}_\mu\bar{\nabla}_\nu \varphi=0,
\end{equation}
where $\delta R_{\mu\nu}(h)$ is given by (\ref{ricc-t}). Its trace
equation is satisfied automatically when one uses (\ref{lriccicon})
and (\ref{linsca}). The metric perturbation $h_{\mu\nu}$ is
classified depending on the transformation properties under parity,
namely odd (axial) and even (polar). Using the Regge-Wheeler
gauge~\cite{Regge:1957td}, and Zerilli gauge~\cite{Zeri}, one
obtains two distinct perturbations : odd and even perturbations. For
odd parity, one has with two off-diagonal components $h_0$ and $h_1$
\begin{eqnarray}
h^o_{\mu\nu}=\left(
\begin{array}{cccc}
0 & 0 & 0 & h_0(r) \cr 0 & 0 & 0 & h_1(r) \cr 0 & 0 & 0 & 0 \cr
h_0(r) & h_1(r) & 0 & 0
\end{array}
\right) e^{-ikt}\sin\Theta\frac{dp_{l}}{d\Theta} \,, \label{oddp}
\end{eqnarray}
while for even parity, the metric tensor takes the form with four
components $H_0,~H_1,~H_2,$ and $K$ as
\begin{eqnarray}
h^e_{\mu\nu}=\left(
\begin{array}{cccc}
H_0(r) f & H_1(r) & 0 & 0 \cr H_1(r) & H_2(r)f^{-1} & 0 & 0 \cr 0 &
0 & r^2 K(r) & 0 \cr 0 & 0 & 0 & r^2\sin^2\Theta K(r)
\end{array}
\right) e^{-ikt}p_{l} \,, \label{evenpt}
\end{eqnarray}
where $p_l$ is Legendre polynomial with angular momentum $l$ and $f$
is the metric function  given by (\ref{num}). Also the scalar
 perturbation is
 \begin{equation}
\varphi(t,r,\Theta,\Phi)=e^{-ik t}\frac{\psi(r)}{r}
 Y_{lm}(\Theta,\Phi).
\end{equation}
 For the odd-parity
perturbation, its linearized equation takes a simple form as
\begin{equation}
\delta R_{\mu\nu}(h)=0,
\end{equation}
which shows that the odd-perturbation is stable, since this is the
same equation as the Eq. (\ref{cur1}). For the even-perturbation
(\ref{evenpt}), however, we have to use the linearized equation
(\ref{jordan-eq})
 because the scalar field $\psi(r)$
 contributes to making an even mode $\hat{M}$ together with $H_0,~H_1,~H_2,$
 and $K$. For example, one has a relation of $H_2=H_0/f^2-2\psi/rf$.
In this case, we have  the Zerilli's equation~\cite{Zeri}
\begin{equation}
\frac{d^2\hat{M}}{dr^{*2}}+\Big[k^2-V_{\rm Z}\Big]\hat{M}=0,
\end{equation}
where $\hat{M}$ and the Zerilli potential are given
by~\cite{Kwon:1986dw,Myung:2011ih,Moon:2011fw}
\begin{eqnarray}
\hat{M}&=&\frac{1}{pq-h}\Big[p(K+\frac{\psi}{r})-\frac{H_1}{k}\Big],\\
V_{\rm
Z}(r)&=&\Big(1-\frac{r_0}{r}\Big)\Bigg[\frac{2\lambda^2(\lambda+1)r^3+3\lambda^2r_0r^2+9\lambda
r_0^2 r/2+9r_0^3/4} {r^3(\lambda r+3r_0/2)^2}\Bigg]
\end{eqnarray}
with
\begin{equation}
\lambda=\frac{1}{2}(l-1)(l+2).
\end{equation}
The Zerilli potential $V_{Z}$ is always positive for whole range of
$-\infty <r^*<\infty$  and $l\ge 2$. Also, it is  a barrier-type
localized around $r^*=0$  which implies that the even-perturbation
is stable, even though the scalar is coupled to the even-parity
perturbations.

The above all statements show clearly that the Schwarzschild black
hole is stable against metric and scalar perturbations (3=$2+1$ DOF)
in the Brans-Dicke theory.

Now let us  go back to the linearized massive equation
(\ref{silin-eq}) in the BDW gravity.    It might be difficult to
solve (\ref{silin-eq}) directly because it is a coupled second-order
equation for $\delta R_{\mu\nu}$ and $\varphi$ ($6=5+1$ DOF).
Surely, this is a nontrivial task.  Curiously, however, the equation
(\ref{silin-eq}) could be rewritten in terms of $\delta\tilde{
R}_{\mu\nu}=\delta R_{\mu\nu}-\bar{\nabla}_\mu\bar{\nabla}_\nu
\varphi$ as
\begin{equation} \label{tildeR-eq} \bar{\nabla}^2\delta
\tilde{R}_{\mu\nu}+2\bar{R}_{\rho\mu\sigma\nu}\delta
\tilde{R}^{\rho\sigma}=\tilde{m}^2\delta \tilde{R}_{\mu\nu},
\end{equation}
where $\tilde{m}^2=m^2\bar{\phi}$ and we used
\begin{eqnarray}
\Delta_L \delta \tilde{R}_{\mu\nu}=\Delta_L \delta R_{\mu\nu}.
\end{eqnarray}
Explicitly, we have the following relation
\begin{eqnarray}
\Delta_L (\bar{\nabla}_\mu\bar{\nabla}_\nu \varphi)
&=&\frac{1}{2}\Delta_L
\Big(\bar{\nabla}_\mu\bar{\nabla}_\nu+\bar{\nabla}_\nu\bar{\nabla}_\mu\Big)\varphi
\nonumber
\\
 &=&-\frac{1}{2}\Big(\bar{\nabla}_\mu\bar{\nabla}_\nu+\bar{\nabla}_\mu\bar{\nabla}_\nu\Big)\bar{\nabla}^2\varphi=0,
 \label{relation}
\end{eqnarray}
 where in the second line, we used the scalar equation (\ref{linsca}). See Appendix for a detailed
 proof of the relation (\ref{relation}).

Obtaining the linearized equation (\ref{tildeR-eq}) is our main
result for carrying out the stability analysis of the Schwarzschild
black hole in the Jordan frame.   It is important to note that the
equation (\ref{tildeR-eq}) actually describes the massive spin-2
field (5 DOF) propagating around the Schwarzschild black hole,
because $\delta\tilde{R}_{\mu\nu}$ satisfies the transverse and
traceless gauge condition:
\begin{eqnarray}
\bar\nabla^{\mu}\delta\tilde{R}_{\mu\nu}~
=~\bar\nabla_{\nu}\delta\tilde{R}~=~0,
\end{eqnarray}
where the Eq. (\ref{linsca}) was used. Note also that the linearized
equation (\ref{tildeR-eq}) is exactly the same as the one
(\ref{eflin-eq}) obtained in the Einstein frame, when replacing
\begin{eqnarray}
\delta\tilde{R}_{\mu\nu}\to\delta\hat{R}_{\mu\nu}
,~~~~~\bar{R}_{\rho\mu\sigma\nu}\to\bar{\hat{R}}_{\rho\mu\sigma\nu}
,~~~~~\tilde{m}^2\to m^2.
\end{eqnarray}
Therefore, the corresponding result can be seen that the unstable
modes for the Schwarzschild black hole in the Jordan frame are given
by the region [see Fig.1]:
\begin{equation} \label{unst-con1} 0<\tilde{m}<\frac{{\cal
O}(1)}{r_0}.
\end{equation}
This states  clearly that the instability of black hole  in the
Jordan and Einstein frames are equivalent.

\section{Discussions}

It was well known that the Schwarzschild black hole is stable
against metric and scalar perturbations in the Brans-Dicke theory
and Einstein gravity. We note that the metric perturbation
$h_{\mu\nu}$ was used to describe a massless spin-2 graviton,
whereas the linearized Ricci tensors of  $\delta \hat{ R}_{\mu\nu}$
and $\delta \tilde{ R}_{\mu\nu}$ were taken to describe (boosted-up)
massive spin-2 graviton to avoid ghost states.  In this work, we
have found unstable $s$-mode from the massive spin-2 graviton
described by the linearized Ricci tensors in the ESW gravity and BDW
gravities.  This implies that the (in)stability of black holes  does
not depend on the frame.

 Let us question what it means that the instability of Schwarzschild
 black hole is given by the $s$-mode of massive spin-2 graviton.
The Schwarzschild black hole stands out among all possible solutions
of Einstein gravity as the only static regular solution to the
vacuum Einstein equation in asymptotically flat spacetimes. The
Schwarzschild solution also solves many other equations of STT,
$f(R)$ gravity, and Chern-Simons gravity including the BDW and ESW
gravity theories. These properties are consistent with various
no-hair proofs which states that the Schwarzschild black hole could
not support regular scalar, nor other fields.  The stability of the
black hole implies that the black hole is really existed as a truly
solution in the Einstein and Brans-Dicke gravity. Hence, the
presence of unstable $s$-mode around the black hole in the BDW and
ESW gravity theories indicates that these massive gravity theories
could not accommodate the static black holes. Naively, the black
holes decay to something and, the final state may be a  spherically
symmetric black hole~\cite{Brito:2013wya,Brito:2013xaa}.
Alternatively, it implies that there is no propagating massive
graviton around the stable Schwarzschild black hole because the
massive graviton is unstable. Therefore, it may happen that the
massive graviton  decays to other fields around the small stable
black holes.

\vspace{3cm}

{\bf Acknowledgement}

\vspace{0.25cm}
 This work was supported by the National
Research Foundation of Korea (NRF) grant funded by the Korea
government (MEST) (No.2012-R1A1A2A10040499).

\section*{Appendix: Proof of the relation (\ref{relation})}
The Lichnerowicz operator acting on a
symmetric second rank tensor
 $M_{ab}$ in the Schwarzschild background is given by
 \cite{Gibbons:2002pq}
 \begin{eqnarray}\label{mani}
\Delta_{\rm
L}M_{ab}~=~2\bar{R}^{c}_{~abd}M^{d}_{~c}-\bar{\nabla}^2M_{ab},
\end{eqnarray}
where
$M_{ab}\equiv\bar\nabla_{(a}V_{b)}\equiv\bar\nabla_{(a}\bar\nabla_{b)}\varphi$.\\
We note that $\bar{\nabla}^2M_{ab}$ can be arranged into the form:
\begin{eqnarray}
\bar{\nabla}^2M_{ab}&=&\frac{1}{2}\bar{\nabla}^2\bar{\nabla}_aV_b+(a\leftrightarrow
b)\nonumber\\
&=&\frac{1}{2}[\bar{\nabla}_c\bar{\nabla}^c,\bar{\nabla}_a]V_b
+\frac{1}{2}\bar{\nabla}_a\bar{\nabla}^2V_b+(a\leftrightarrow
b)\nonumber\\
&=& \frac{1}{2}\bar{\nabla}_c[\bar{\nabla}^c,\bar{\nabla}_a]V_b
+\frac{1}{2}[\bar{\nabla}_c,\bar{\nabla}_a]\bar{\nabla}^cV_b
+\frac{1}{2}\bar{\nabla}_a\bar{\nabla}^2V_b+(a\leftrightarrow
b)\nonumber\\
&=& \frac{1}{2}\bar{\nabla}_c\Big(\bar{R}_{b~~~a}^{~\lambda
c}V_{\lambda}\Big)
+\frac{1}{2}\bar{R}_{b~~ca}^{~\lambda}\bar{\nabla}^{c}V_{\lambda}
+\frac{1}{2}\bar{\nabla}_a\bar{\nabla}^2V_b+(a\leftrightarrow
b)\nonumber\\
&=& \frac{1}{2}\bar{\nabla}_c\bar{R}_{b~~~a}^{~\lambda
c}V_{\lambda}+\bar{R}_{b~~ca}^{~\lambda}\bar{\nabla}^{c}V_{\lambda}
+\frac{1}{2}\bar{\nabla}_a\bar{\nabla}^2V_b+(a\leftrightarrow b)
\nonumber\\
&=&\frac{1}{2}\bar\nabla_{c}\bar{R}^{c}_{~ab\lambda}V^{\lambda}
+\bar{R}^{c}_{~~ab\lambda}\bar{\nabla}_{c}V^{\lambda}+
\frac{1}{2}\bar{\nabla}_a\bar{\nabla}^2V_b+(a\leftrightarrow b)
\nonumber\\
&=&\bar\nabla_{c}\bar{R}^{c}_{~(ab)\lambda}V^{\lambda}
+2\bar{R}^{c}_{~ab\lambda}M^{\lambda}_{~c}
+\bar\nabla_{(a}\bar{\nabla}^2V_{b)}.\label{aaa1}
\end{eqnarray}
Substituting (\ref{aaa1}) into Eq. (\ref{mani}) leads to
\begin{eqnarray}\label{fin}
\Delta_{\rm
L}M_{ab}~=~-\bar\nabla_{c}\bar{R}^{c}_{~(ab)\lambda}V^{\lambda}
-\bar\nabla_{(a}\bar{\nabla}^2V_{b)}.
\end{eqnarray}
From the Bianchi idenity, the first term of the r.h.s. in Eq.
(\ref{fin}) vanishes:
\begin{eqnarray}
&&\bar{\nabla}_{[\mu}\bar{R}_{ab]cd}~=~0\nonumber\\
 \Rightarrow&&\bar{\nabla}_{\mu}\bar{R}_{abcd}
 +\bar{\nabla}_{a}\bar{R}_{b\mu cd}
 +\bar{\nabla}_{b}\bar{R}_{\mu acd}~=~0~~(\times~ \bar{g}^{ac})\nonumber\\
 \Rightarrow&&\bar{\nabla}_{\mu}\bar{R}_{bd}+\bar{\nabla}^{c}\bar{R}_{b\mu
 cd}-\bar{\nabla}_{b}\bar{R}_{\mu d}~=~0\label{r1}\\
\Rightarrow&&\bar{\nabla}_{c}\bar{R}^{c}_{~db\mu}~=~0.\label{r2}
\end{eqnarray}
The second term of the r.h.s. in Eq. (\ref{fin}) can be re-written
as follows:
\begin{eqnarray}
\bar{\nabla}^2V_{b}~=~\bar{\nabla}^2\bar{\nabla}_{b}\varphi
&=&[\bar{\nabla}_c\bar{\nabla}^c,\bar{\nabla}_b]\varphi
+\bar{\nabla}_b\bar{\nabla}^2\varphi\nonumber\\
&=&\bar{\nabla}_{c}[\bar{\nabla}^{c},\bar{\nabla}_{b}]\varphi
+[\bar{\nabla}_{c},\bar{\nabla}_{b}]\bar{\nabla}^{c}\varphi
+\bar{\nabla}_b\bar{\nabla}^2\varphi\nonumber\\
 &=&\bar{R}_{\lambda b}\bar{\nabla}^{\lambda}\varphi
+\bar{\nabla}_b\bar{\nabla}^2\varphi\label{phi1}\\
&=&\bar{\nabla}_b\bar{\nabla}^2\varphi\nonumber\\
 &&\hspace*{-6em}
\Rightarrow~\bar\nabla_{a}\bar{\nabla}^2V_{b}
~=~\bar\nabla_{a}\bar{\nabla}^2\bar{\nabla}_b\varphi~=~
\bar\nabla_{a}\bar{\nabla}_b\bar{\nabla}^2\varphi.\label{phi2}
\end{eqnarray}
Note that in Eqs. (\ref{r1}) and (\ref{phi1}), we used the Ricci
flat condition of $\bar{R}_{ab}=0$, which shows  the Schwarzschild
background. Plugging the Eqs. (\ref{r2}) and (\ref{phi2}) into Eq.
(\ref{fin}), we finally get
\begin{eqnarray}
\Delta_{\rm L}M_{ab}~=~\Delta_{\rm
L}\bar{\nabla}_{(a}\bar{\nabla}_{b)}\varphi
~=~-\bar{\nabla}_{(a}\bar{\nabla}_{b)}\bar{\nabla}^2\varphi.\nonumber
\end{eqnarray}

\end{document}